\documentclass{moriond}

\usepackage{graphicx} 
\usepackage{amsmath}  
\usepackage{amssymb} 
\usepackage{cleveref}

\def\be{\begin{equation}}
\def\ee{\end{equation}}
\def\bea{\begin{eqnarray}}
\def\eea{\end{eqnarray}}

\newcommand{\Photo}{\includegraphics[height=35mm]{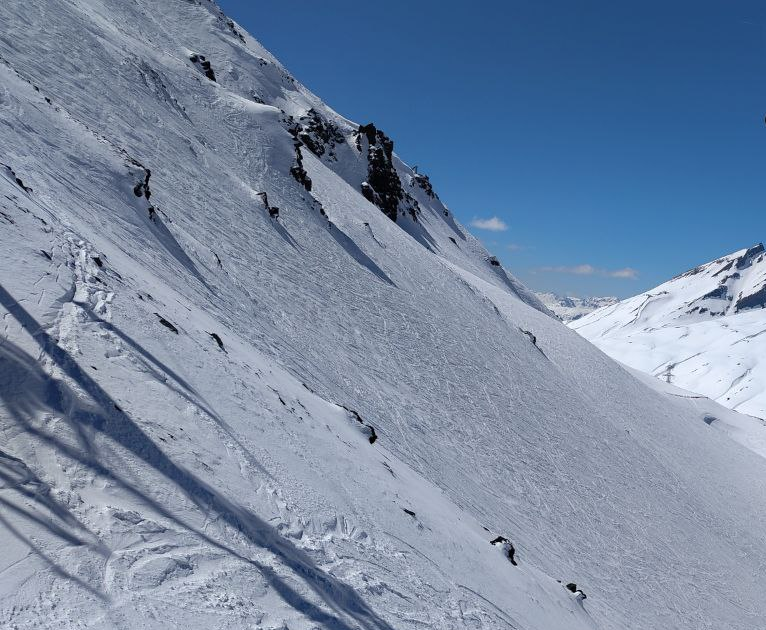}}

\Crefname{figure}{Fig.}{Fig.}

\begin{document}
\vspace*{4cm}
\title{Constraining primordial curvature perturbations with present and future GW detectors}

\author{Mauro Pieroni}

\address{Instituto de Estructura de la Materia, CSIC, Calle de Serrano 121, 28006 Madrid, Spain\\
Theoretical Physics Department, CERN, 1 Esplanade des Particules, Geneva 23, CH-1211, Switzerland \\ 
Email: mauro.pieroni@csic.es}

\maketitle
\abstracts{
    Primordial scalar curvature perturbations ($\zeta$), typically probed on large cosmological scales via CMB and LSS observations, can be significantly enhanced on smaller scales by various early Universe mechanisms, for instance, non-minimal inflationary models. While decoupled at linear order, scalar and tensor perturbations, i.e., Gravitational Waves (GWs), interact at second order. As a consequence, an enhanced primordial scalar power spectrum $P_\zeta(k)$ can source a sizable stochastic GW background (SGWB). In this proceedings, we briefly review the generation mechanism of such signals, typically referred to as scalar-induced GWs (SIGWs), and discuss the prospects of measuring them with present and future Pulsar Timing Arrays datasets and future GW observatories like the Laser Interferometer Space Antenna LISA.  }

\section{Introduction}

At CMB scales ($k \sim 10^{-4} - 10^{-1} \, {\rm Mpc}^{-1}$), primordial scalar perturbations, $\zeta$, are observed to have an amplitude $\sim 10^{-5}$. However, on smaller scales they remain rather unconstrained, and they could, in principle, possess a much larger amplitude. At second order in perturbation theory, scalar perturbations source tensor perturbations~\cite{Tomita75,Matarrese93,Domenech21Review}, typically referred to as Scalar Induced Gravitational Waves (SIGWs). Thus, GW detectors can be used to probe $P_\zeta(k)$ on much smaller scales compared to other cosmological surveys, which are otherwise inaccessible.

Several Pulsar Timing Array (PTA) collaborations such as NANOGrav~\cite{NANOGrav23} and EPTA/InPTA~\cite{EPTA23}, have recently reported evidence for a common-spectrum process across their monitored pulsars, consistent with an isotropic Stochastic GW Background (SGWB) in the nHz frequency band ($k \sim 10^{6} - 10^{8} \, {\rm Mpc}^{-1}$). Beyond the leading astrophysical interpretation, cosmological sources, including SIGWs, remain viable explanations for this signal~\cite{NANOGravCosmo23}. Moreover, in the next decade, the Laser Interferometer Space Antenna (LISA)~\cite{LISA:2017pwj} will survey the mHz frequency band, corresponding to much smaller scales ($k \sim 10^{11} - 10^{14} \, {\rm Mpc}^{-1}$) than those probed by PTAs or the CMB. After briefly reviewing the SIGW generation mechanism, in this proceeding, we discuss the prospects of constraining scalar perturbations with present and future PTA experiments and with LISA.

\section{Scalar-induced Gravitational Waves}

The spectrum of tensor perturbations $P_h(k)$ generated at second order by scalar perturbations can be described by the integral:
\begin{equation}
    \hspace*{-0.3cm} P_h(k, \eta) = 4 \int_0^{\infty} \textrm{d} t \int_{-1}^{1} \textrm{d} s \; \frac{t^2(1-s^2)^2}{(1+t)^2} \;  I^2(k, t, s, \eta) \;  P_\zeta(k u)  \;  P_\zeta(k v),
    \label{eq:Ph_integral}
\end{equation}
where $\eta$ is the conformal time, $t$ and $s$ are auxiliary variables to compute the double integral over the two scalar momenta, $u = (1+t-s)/2$, $v = (1+t+s)/2$, and $I(k, t, s, \eta)$ is a kernel function depending on the cosmological transfer functions, which describe the time evolution of the perturbations.  

Since the $P_h$ is proportional to $P_\zeta^2$, which in the minimal $\Lambda$CDM model is nearly scale-invariant and of the order of $10^{-9}$, the resulting SIGW spectrum is typically very small. However, if $P_\zeta(k)$ is significantly enhanced at small scales, the resulting SIGW signal can be large enough to be measured with GW detectors. Many early Universe mechanisms, including several inflationary scenarios (e.g., ultra slow-roll, non-standard kinetic terms, and multi-field models), can predict such an enhancement of $P_\zeta(k)$. To remain agnostic about the specific model, we can parametrize the enhancement of $P_\zeta(k)$ with a lognormal peak:
\begin{equation}
    P_\zeta(k) = \frac{A}{\sqrt{2 \pi} \Delta} \exp\left(-\frac{1}{2} \left(\frac{\ln(k/k_0)}{\Delta}\right)^2\right) \; ,
    \label{eq:lognormal}
\end{equation}
where $A$ is the amplitude of the peak, $k_0$ is the peak scale, and $\Delta$ is the width of the peak.

While, in general, the integral in~\cref{eq:Ph_integral} should be computed numerically (see, e.g., \href{https://github.com/jonaselgammal/SIGWAY}{\texttt{SIGWAY}}, for a numerical calculation tools developed by the LISA Cosmology Working Group~\cite{LISACosmologyWorkingGroup:2025vdz}), the resulting GW energy density spectrum $\Omega_{\rm GW}(f) \propto k^2 P_h(k)$ exhibits some characteristic features~\cite{LISACosmologyWorkingGroup:2025vdz}:\\ 
- For narrow $P_\zeta$, the SIGW spectrum exhibits a double peak structure.\\
- In the infrared (i.e., below the peak scale), $P_h$ exhibits a universal $f^3$ scaling due to causality.\\
- In the ultraviolet (i.e., above the peak scale), $P_h$ is nearly proportional to $P_\zeta^2$.\\
Moreover, some analytic approximations for the SIGW spectrum for a lognormal $P_\zeta$ can be found in the literature~\cite{Pi:2020otn}.

\section{Constraints/prospects from Pulsar Timing Arrays}

The most common astrophysical interpretation for the current PTA measurements, is to attribute the signal to a population of merging supermassive black hole binaries (SMBHBs)~\cite{NANOGrav23}, approximately a power law (with $\alpha_{\rm PL}$ denoting the log-amplitude and $n_{\rm T}$ the tilt). However, the near infrared part of the SIGW spectrum provides a good fit to the observed signal~\cite{NANOGravCosmo23}. Current data, cannot give a conclusive answer on the nature of the observed signal. Thus, it is interesting to forecast the improvements with longer observation times $T_{\rm obs}$ and with future PTA experiments, such as the Square Kilometre Array (SKA)~\cite{Janssen:2014dka}. We have performed our forecasts~\cite{Cecchini25}, with the open-source \href{https://github.com/Mauropieroni/fastPTA/}{\texttt{fastPTA}}~\cite{Babak:2024yhu} tool, and we have explored two different scenarios:

\textbf{Scenario 1:} \emph{Assuming PTAs are currently observing SIGWs}, we have investigated how the constraints improve with future configuration. We found~\cite{Cecchini25} that for an SKA-like detector with $N_p = 70$ pulsars, constraints improve significantly, with relative errors on the three parameters of the lognormal peak ($A_\zeta, \Delta, k_*$) of the order of $10\%$, scaling as $\sqrt{70 / N_p}$ with $N_p$.

\begin{figure}[htb]
    \centering
    \includegraphics[width=0.5\linewidth]{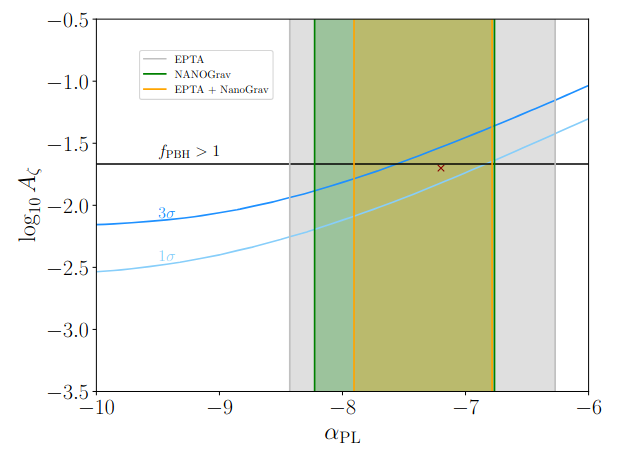}
    \includegraphics[width=0.45\linewidth]{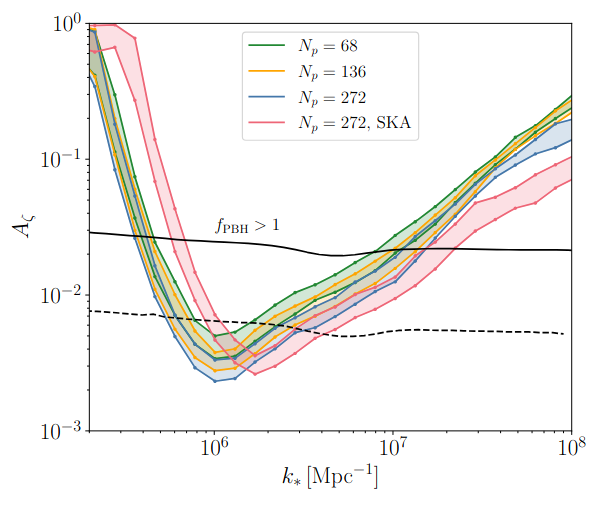}
    \caption{Left: detectability level for $A_\zeta$. The vertical shaded areas represent the current PTA bounds (see legend) on $\alpha_{\rm PL}$, the black horizontal line represents the limit imposed by Primordial Black Hole (PBH) overproduction. Right: Projected 95-99.7\% CL upper limits on $A_\zeta$ for EPTA with $T_{\rm obs}= 16.03$yrs and SKA with $T_{\rm obs}= 10$yrs varying $N_p$ (see legend). Solid/dashed black lines denotes PBH overproduction (threshold statistics/peak theory).}
    \label{fig:pta_forecasts}
\end{figure}

\textbf{Scenario 2:} \emph{Assuming the current signal is mostly due to SMBHBs with a subdominant SIGW component}, we have forecasted constraints (detection)/upper limits (no detection) on the SIGW parameters. \Cref{fig:pta_forecasts} (left) shows the detectability level for $A_\zeta$ for different values of $\alpha_{\rm PL}$, marginalized over all other parameters (injected values $n_{\rm T} = 2, \log_{10}\Delta = -0.3, \log_{10}(f_*/{\rm Hz}) = -7.8)$ compared to experimental constraints.~\Cref{fig:pta_forecasts} (right) shows the projected marginalized 95-99.7\% CL upper limits on the SIGW signal (fit using SMBHBs + SIGW on data contain SMBHBs only) in the $(A_\zeta,k_*)$ plane for several SKA configurations. For more details, see~\cite{Cecchini25}.

\begin{figure}[htb]
    \centering
    \includegraphics[width=0.95\linewidth]{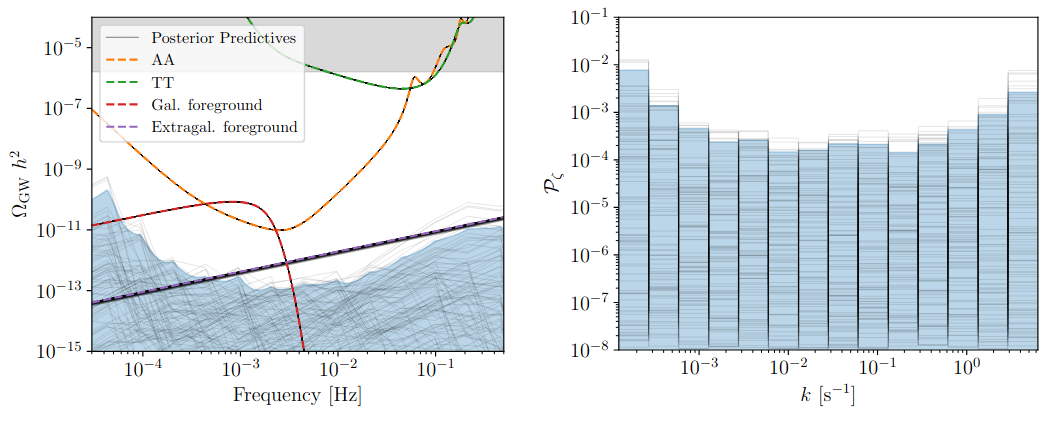} 
    \caption{Left: forecast LISA upper limit on $\Omega_{\rm GW} h^2$ (cyan band) compared with the astrophysical foregrounds and detector sensitivities for the different data channels (see legend). Right: bin-by-bin upper
    limits on $P_\zeta$.}
    \label{fig:lisa_upper_limit}
\end{figure}

\section{Prospects with LISA}

In the LISA band, corresponding to much smaller scales compared to CMB and PTAs, the only constraints on $\Omega_{\rm GW}$ and on $P_\zeta(k)$, come from PBH overproduction and from the BBN bound, which are not very stringent. Thus, there is a large parameter space for SIGWs that LISA will probe/constrain. In~\cite{LISACosmologyWorkingGroup:2025vdz}, a collaborative project within the LISA Cosmology Working Group (CosWG), we have thoroughly investigated the prospects for LISA to detect and characterize SIGWs. Some of the main the results of this work are:\\
- Systematic scans of the parameter space (again assuming a lognormal $P_\zeta$) of the SIGW signals, performed with Fisher Information Matrix (FIM) formalism and validated with full Bayesian analysis (via Monte Carlo techniques). We have found that $(A_{\zeta}, \Delta, k_* )$ can be measured with accuracy below $1\%$ (depending on the injected signal), for Signal-to-Noise Ratio (SNR) $\gtrsim 1000$. \\
- Assuming no SGWB signal is measured, and using a binned parameterization for $P_{\zeta}$ (for details, see~\cite{LISACosmologyWorkingGroup:2025vdz}), we have shown that LISA will constrain $P_\zeta$ to the level of $A_{\zeta} \lesssim 10^{-4}$, see~\cref{fig:lisa_upper_limit}. \\
- By performing an end-to-end analysis of ultra-slow roll inflationary models, we have shown that, for models generating a sufficienly large signal, LISA will constrain the shape of the inflationary potential to the order of $10\%$ in a region where it is currently unconstrained.\\
- LISA could potentially probe the presence of primordial non-Gaussianities on small scales potentially reaching $\mathcal{O}(1\%)$ sensitivity for $f_{\rm NL} \sim 10$.

\section{Conclusions and Outlook}

SIGWs offer a unique probe of $P_\zeta$ on scales far beyond the reach of CMB and LSS observations. Current PTA data provide intriguing hints of an SGWB of uncertain origin in the nHz frequency band. Since the constraining power will improve significantly with longer observation times/better instrumental configurations/monitoring more pulsars, future PTA observations will better measure and characterize the SGWB spectrum, which is crucial to distinguish SMBHB from other signals, such as SIGWs. Thus, in the near future, PTA experiments have the potential to either accurately constrain the SIGW model parameters, or set the most stringent upper limit on $P_\zeta$ for comoving wavevector  in the range $k \sim 10^{11} - 10^{14} \, {\rm Mpc}^{-1}$. Similarly,  on even smaller scales, in the case of detection, LISA should provide tight constraints on the SIGW model parameters, or, alternatively, set an upper limit on $P_\zeta$ to the level $A_{\zeta} \lesssim 10^{-4}$.

The complementarity between GW detectors and other cosmological probes will enable us to constrain primordial curvature perturbations on a wide range of scales. However, fully exploiting the potential of future experimental observations requires accurate theoretical predictions and robust data analysis methods, posing several interesting challenges for the upcoming years.

\section*{Acknowledgments}
This proceeding is based on works with Gabriele Franciolini and Chiara Cecchini~\cite{Cecchini25} and with members of the LISA CosWG~\cite{LISACosmologyWorkingGroup:2025vdz}. The work of MP is funded by a Contrato de Atracción de Talento de la Comunidad de Madrid (Spain), with number 2024-T1TEC-31343. 

\section*{References}
\bibliography{bibliography} 

\end{document}